\documentclass[twocolumn,showpacs,preprintnumbers,amsmath,amssymb]{revtex4}
\usepackage{graphicx}

\begin{document}

\title{Effective mass analysis of Bose-Einstein condensates in optical lattices --- \\
stabilization and levitation}

\author{H. Pu$^1$, L. O. Baksmaty$^2$, W. Zhang$^1$, N. P. Bigelow$^2$
 and P. Meystre$^1$}
\affiliation{$^1$Optical Sciences Center, The University of
Arizona, Tucson, AZ 85721 \\
$^2$Dept. of Physics and Astronomy, and Laboratory for Laser
Energetics, University of Rochester, Rochester, NY 14620}
\date{\today}

\begin{abstract}
We investigate the time evolution of a Bose-Einstein condensate in
a periodic optical potential. Using an effective mass formalism,
we study the equation of motion for the envelope function
modulating the Bloch states of the lattice potential. In
particular, we show how the negative effective mass affects the
dynamics of the condensate.
\end{abstract}
 \maketitle

\section{Introduction}

The dynamics of classical and quantum systems in periodic
potentials is a central paradigm of physics, finding applications
from condensed matter physics to optics, and from nonlinear
dynamics to atomic and to plasma physics. In recent years, the
experimental and theoretical study of quantum-degenerate atomic
systems in periodic potentials has opened up new avenues of
investigation. In particular, Bose-Einstein condensates are a
macroscopic quantum system that is amenable to exquisite
experimental control. As a result, many phenomena studied in solid
state systems can be re-examined in a more direct and dramatic
fashion. Even more importantly perhaps, it is now possible to
realize experimentally model systems that had previously been the
object of considerable theoretical studies, but were all but
impossible to test experimentally. One particularly beautiful
example is the experimental realization of the Hubbard model
leading to the demonstration of the superfluid to Mott insulator
transition in a Bose condensate of $^{87}$Rb atoms~\cite{mott}.

The first experiments involving the dynamics of Bose-Einstein
condensates (BEC) in periodic potentials were carried out by
Anderson and Kasevich, who used this approach to demonstrate a
mode-locked atom laser \cite{modelock} and observe atomic
Josephson oscillations \cite{modelock,josephson}. The list of
phenomena that were subsequently experimentally demonstrated
and/or theoretically investigated includes the generation of
atomic number squeezing \cite{squeez}, the observation of the
superfluid-Mott insulator phase transition, the generation of
discrete \cite{dis} or gap solitons \cite{gap}, the prediction of
modulational instabilities \cite{stability} and superfluid flow
\cite{flow}, the observation of Bloch oscillations
\cite{blochosc}, the analysis and observation of coherent
acceleration \cite{acceleration}, studies of magnetism \cite{mag},
etc.

It is well known that a particle confined to an infinite periodic
potential and acted upon by an external force behaves as if
possessing an effective mass that can be substantially different
from its true mass, and may even take negative values. In
particular, it is this property that is at the core of proposals
to generate bright matter-wave solitons in BECs with repulsive
interactions \cite{gap}. The purpose of the present paper is to
extend these studies by analyzing the temporal evolution of a
condensate in the periodic potential provided by an optical
lattice. The initial state of the condensate is chosen to be a
(approximate) Bloch state modulated by a slow-varying Gaussian
envelope, and we compare two situations where the Bloch state is
either associated with a positive or a negative effective mass. We
show that in situations where the dynamics of the condensate is
well approximated by a negative effective mass, the periodic
potential can lead to the stabilization of an otherwise unstable
condensate. We also demonstrate theoretically the levitation of
condensates of negative effective masses. These studies further
allow us to determine the impact of self-interactions and of
finite condensate widths on the general usefulness of the
effective mass concept.

The paper is organized as follows. Section II gives a brief review
of the linear problem of a particle inside a periodic sinusoidal
potential and introduces the concept of effective mass. Section
III applies and extends these ideas to the dynamics of a BEC in an
optical lattice. We discuss the equation of motion for the slowly
varying condensate envelope function, from which we can gain
useful physical insights into the dynamics of the system. A full
numerical solution is presented in Sec.~IV, which demonstrates in
particular the stabilization and the levitation of a condensate of
negative effective mass, in agreement with the prediction of
Sec.~III. Finally, Section V presents concluding remarks on the
usefulness of the effective mass concept and an outlook.

\section{Linear properties of infinite periodic potential}

\begin{figure}
\includegraphics*[width=7cm,height= 8.5 cm]{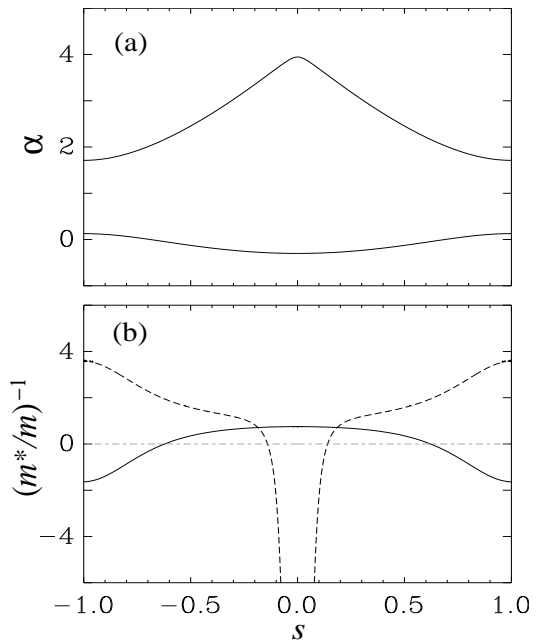}
\caption{(a) Dimensionless energy $\alpha=b-h^2/2$ and (b) scaled
reciprocal effective mass $m/m^*$ for the first two bands as a
function of the normalized quasi-momentum $s$. In (b), the solid
and dashed lines refer to the first and second band, respectively.
For this calculation, $h^2=3.2$. } \label{fig1}
\end{figure}

In this section, we briefly review important aspects of the linear
problem of a particle of mass $m$ inside a one-dimensional
infinite periodic potential of the form
\begin{equation}
V(x)=V_0 \cos^2 (k_0 x), \label{potential}
\end{equation}
with corresponding time-independent Schr\"{o}dinger equation
\begin{equation}
\left[ -\frac{\hbar^2}{2m} \frac{d^2}{dx^2} + V_0 \cos^2 (k_0 x)
\right] \phi(x) = E \phi(x). \label{se1}
\end{equation}
We proceed by introducing the dimensionless quantities
\begin{equation}
\theta = k_0 x;\;\;\;b=\frac{2mE}{\hbar^2
k_0^2};\;\;\;h^2=\frac{2mV_0}{\hbar^2 k_0^2},\label{para}
\end{equation}
in terms of which Eq. (\ref{se1}) becomes
\begin{equation}
\frac{d^2 \phi}{d \theta^2} +\left( b-h^2 \cos^2 \theta \right)
\phi =0. \label{se2}
\end{equation}
Eq.~(\ref{se2}) has the form of Mathieu's equation, whose
solutions are well known \cite{mathieu}. As a warm-up we sketch
out the main features of its solutions which, according to Bloch's
theorem, must be of the form
\[\phi_{n,s}(\theta)= e^{is\theta} F_n(\theta),\]
where $s$ is real and arbitrary (in standard textbook language,
$s=k/k_0$ where $\hbar k$ is the quasi-momentum). In our scaled
units, the periodic functions $F_n(\theta)$ have a period of
$\pi$. The energy spectrum associated with the periodic potential
$V(\theta)$ exhibits a band structure familiar from solid state
physics. Each value of $s$ gives a discrete spectrum whose
structure is periodic with respect to $s$. This property allows
one to restrict the discussion to the first Brillouin zone, $-1 <
s \leq 1$. As $s$ is increased within that zone, the energy levels
trace out curves that are restricted to a small energy band. We
restrict our discussion to the first two bands, as depicted in
Fig.~\ref{fig1}(a). Note that although the probability
distribution $|\phi|^2$ is a periodic function of $\theta$, $\phi$
itself is not periodic unless $s$ takes integer values --- e.g.,
at the center or the edges of the Brillouin zone. For a detailed
discussion of the solutions to Eq.~(\ref{se2}), see e.g.
Ref.~\cite{mathieu}.

For $s=0$ or 1, the eigenstates of the Mathieu's equation are
periodic and take the form of standing waves. In particular, at
the edge $s=1$ of the first Brillouin zone, the wave functions
$\phi_{n,s}$ for the first two energy bands, $n=1,2$, can be
expressed as the Fourier series
\begin{equation}
 \phi_{1,s=1}(\theta) =
\sum_{n=0}^{\infty} A_{2n+1} \sin (2n+1)\theta,\label{neg}
\end{equation}
and
\begin{equation}
 \phi_{2,s=1} (\theta) = \sum_{n=0}^{\infty} B_{2n+1}
\cos (2n+1)\theta,\label{pos}
\end{equation}
where
$$G_{2n+1}=k_{2n-1}-\frac{1}{G_{2n-1}},\;\;\;n>1,$$
$$k_n=(4b-2h^2-4n^2)/h^2,$$
$$G_{2n+1}=A_{2n+1}/A_{2n-1},\;\;\;{\rm or}\;\;\;
B_{2n+1}/B_{2n-1},$$
and $A_3 = k_1+1$, $B_3=k_1-1$.

The effective mass, $m^*$, which characterizes the response of the
particle to external perturbations, is defined as
\begin{equation}
m^*= \frac{\hbar^2}{\partial^2 E /\partial
k^2}=\frac{2m}{\partial^2 b/
\partial s^2}.
\end{equation}
Fig.~\ref{fig1}(b) shows the variation of the effective mass of
the first two bands with the quasi-momentum. One observes that it
is near infinite (i.e., $1/m^* =0$) for certain values of the
quasi-momentum $s$, and even becomes negative in certain regions
of the zone. In more than one dimension, $m^*$ acquires a
tensorial character, with elements given by
\[m^*_{\mu \nu}=\frac{\hbar^2}{\partial^2 E /(\partial k_{\mu}
\partial k_{\nu})}\,.\]

In the next section, we apply the effective mass formalism to the
case of a self-interacting Schr{\"o}dinger field described by the
Gross-Pitaevskii equation. Specifically, we show that a BEC
prepared in a Bloch state with negative effective mass behaves as
if the signs of its self-interaction and of the external confining
potential had been reversed.

\section{Effective mass equation}

We consider a BEC in a sinusoidal optical lattice potential. Near
zero temperature, the system is described to an excellent degree
of approximation by a time-dependent Gross-Pitaevskii equation for
the normalized condensate wave function $\psi({\bf r},t)$,
generalized to include the effects of three-body collisions,
\begin{widetext}
\begin{equation}
i \hbar \frac{\partial \psi({\bf r},t)}{\partial t} = \left[
-\frac{\hbar^2}{2m} \nabla^2 + V({\bf r}) + U({\bf r})+N g|\psi|^2
- iN^2 K_3 |\psi|^4 \right] \psi({\bf r},t).\label{gpe}
\end{equation}
\end{widetext}
Here, $V({\bf r})$ is the periodic lattice potential, $U({\bf r})$
is an additional external potential that is taken to be slowly
varying on the scale of the lattice period (e.g., a confining
potential), $N$ is the total number of atoms, $g$ is the strength
of the nonlinear inter-atomic interaction, and $K_3$ gives the
rate of three-body recombination loss \cite{loss}.

We proceed by expanding the condensate wave function $\psi({\bf
r},t)$ on the complete set of Bloch functions $\phi_{n{\bf
k}}({\bf r})$ that are the stationary solutions of the associated
linear Schr\"{o}dinger equation
\begin{equation}
\left[ -\frac{\hbar^2}{2m} \nabla^2 + V({\bf r}) \right]
\phi_{n{\bf k}}({\bf r}) = E_{n{\bf k}}\, \phi_{n{\bf k}}({\bf
r}). \label{bloch}
\end{equation}
The subscripts $n$ and ${\bf k}$ represent the band index and
quasi-momentum, respectively. The Bloch functions satisfy the
orthonormality condition:
\[ \int_{\rm cell}\, d{\bf r}\,\phi_{m{\bf k}}^*({\bf r})\,
\phi_{n{\bf k}'} ({\bf r}) = \frac{\Omega}{(2 \pi)^3}\,
\delta_{mn} \delta_{{\bf k}{\bf k}'}\, ,\] where the integral is
taken over a single period of the lattice of volume $\Omega$. Our
use of the resulting expansion,
\begin{equation}
\psi({\bf r},t)= \sum_{n,{\bf k}}\,A_{n{\bf k}}({\bf r},t)
\phi_{n{\bf k}}({\bf r}) e^{-iE_{n{\bf
k}}t/\hbar}\,,\label{expansion}
\end{equation}
is motivated by the the fact that the Bloch functions can capture
the rapid oscillations of the condensate wave function, while the
slowly varying envelope functions $A_{n{\bf k}}$ will describe the
slow center-of-mass motion of the condensate.

The equations of motion for the amplitudes $A_{n{\bf k}}$ are
found by inserting the expansion (\ref{expansion}) into the
generalized Gross-Pitaevskii equation (\ref{gpe}). The resulting
equations are however of little use if a very large number of
$A_{n{\bf k}}$ is required to accurately describe the dynamics of
the condensate. Here, we consider a more specialized situation
where the matter-wave field is characterized by a central wave
vector ${\bf k}_0$ corresponding to the mean velocity of the
condensate (this would correspond to the carrier wave in
conventional optics). Hence, we expand $\psi({\bf r},t)$ in a way
reminiscent of the slowly varying envelope approximation of
quantum optics as
\begin{equation}
\psi({\bf r},t)= \sum_{n}\,f_{n}({\bf r},t) \phi_{n{\bf k}_0}({\bf
r}) e^{-iE_{n {\bf k}_0}t/\hbar}\,. \label{exp}
\end{equation}
Inserting Eq. (\ref{exp}) into Eq.~(\ref{gpe}) and applying the
effective mass \cite{book} or multiple scales \cite{sipe} methods
yields a set of equations governing the time evolution of the
condensate envelope functions $f_{n}({\bf r},t)$ \cite{steel},
\begin{widetext}
\begin{equation}
i\hbar \left( \frac{\partial f_{n}}{\partial t} + {\bf v}_g \cdot
\nabla f_{n} \right) =\left[ -\frac{\hbar^2}{2m^*_{\mu \nu}}
\frac{\partial^2}{\partial x_{\mu}
\partial x_{\nu}}  + U({\bf r})  + Ng'
|f_{n}|^2 - iN^2 K'_3 |f_{n}|^4 \right] f_{n}. \label{gpe*}
\end{equation}
\end{widetext}
The derivation of Eq.~(\ref{gpe*}), the so-called effective mass
equation, is given in Appendix~\ref{app1}. Here $x_{\mu}$ and
$x_{\nu}$ are cartesian components of ${\bf r}$ and we use an
implicit summation over repeated indices. The velocity
\begin{equation}
 {\bf v}_g=\frac{1}{m} \langle \phi_{n{\bf
k}_0}| \hat{\bf p} | \phi_{n{\bf k}_0} \rangle ,\label{vg}
\end{equation}
is the drift velocity of the $n$-th band contribution to the
condensate envelope. As shown in Appendix~\ref{app1}, it is equal
to the gradient of the energy $E_{n{\bf k}}$ with respect to the
quasi-momentum ${\bf k}$, evaluated at ${\bf k}_0$. We note that
${\bf v}_g$ vanishes at the extreme points of the band, in
particular at the band edges. The coefficients $g'$ and $K'$ give
the renormalized nonlinear interaction strength and three-body
loss rate, respectively. They are given by
\begin{equation}
g'= \frac{(2\pi)^3}{\Omega}g\, \int_{\rm cell}\, d{\bf
r}\,|\phi_{n{\bf k}_0}({\bf r})|^4,
\end{equation}
and
\begin{equation}
 K'_3 = \frac{(2\pi)^3}{\Omega}K_3\, \int_{\rm
cell}\, d{\bf r}\,|\phi_{n{\bf k}_0}({\bf r})|^6
\end{equation}
The lattice potential $V({\bf r})$ does not appear in these
expressions, its effects being already incorporated into the
effective mass $m^*$ which appears in the first term on the
right-hand side of Eq. (\ref{gpe*}) as well as in the drift term
on the left-hand side of that equation.

The effects of a negative mass, $m^*<0$, can be immediately
inferred from the complex conjugate of Eq.~(\ref{gpe*}),
\begin{widetext}
\begin{equation}
i\hbar \left( \frac{\partial f^{\star}_{n}}{\partial t} + {\bf
v}_g \cdot \nabla f^{\star}_{n} \right) =\left[
-\frac{\hbar^2}{2|m^*_{\mu \nu}|} \frac{\partial^2}{\partial
x_{\mu}
\partial x_{\nu}}  - U({\bf r})  - Ng'
|f^{\star}_{n}|^2 -iN^2 K'_3 |f^{\star}_{n}|^4 \right]
f^{\star}_{n}.
\end{equation}
\label{fstar}
\end{widetext}
This equation shows that as compared to the case of a positive
effective mass, the condensate density envelope profile
$|f_n(t)|^2$ evolves now under the influence of an inverted
external potential $-U({\bf r})$ and an inverted nonlinearity
$-g'$. As expected, though, the three-body recombination rate does
not change sign and still represents a loss term. The change in
sign of $g'$ and $U$ leads to a number of consequences. In
particular, past work has shown how this property can be exploited
to launch bright matter-wave solitons in condensates with positive
scattering length \cite{gap} and induce modulational instability
\cite{stability}. The following section explores further
consequences of this property.

\section{Effects of Negative effective mass}

\subsection{Inverted nonlinearity}
We first investigate the effect of an inverted nonlinearity. As is
well known, this will change the two-body interaction from being
repulsive to attractive, and vice versa. Since condensates with
attractive interactions are normally unstable against collapse,
the sign change of $m^*$ is clearly expected to have a significant
effect.

To reduce the scale of the computation, we restrict ourselves to a
two-dimensional system consisting of a pancake-shaped condensate
whose axial motion is frozen to the trap ground state. (We note
that condensates with reduced dimensions have already been
realized in the laboratory \cite{reduce}.) Furthermore, to focus
on the effects of the nonlinearity, we assume that except for the
lattice potential
\begin{equation}
V(x,y)=V_0\,[\cos^2 (k_0 x) + \cos^2 (k_0 y)] ,
\end{equation}
there is no additional external potential in the transverse
dimensions.

\begin{figure}
\includegraphics*[width=7cm,height=11.4cm]{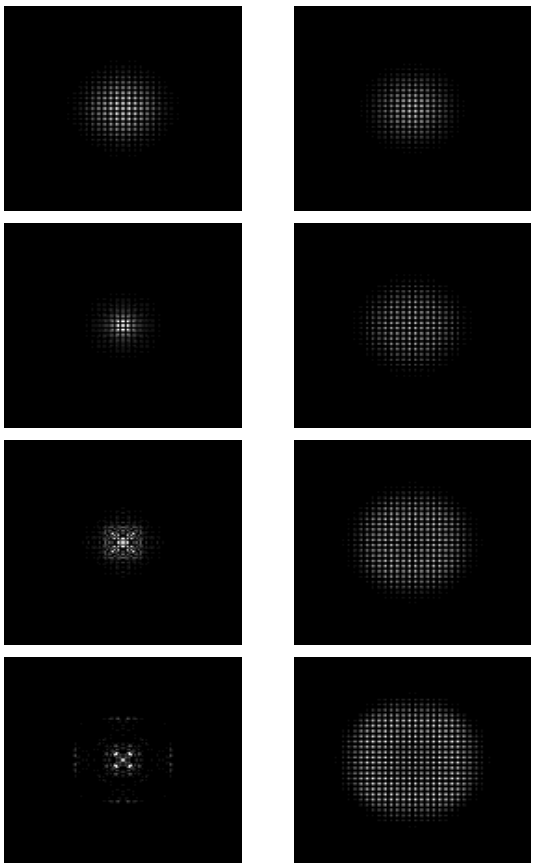}
\caption{Temporal evolution of the density distribution of a BEC
in optical lattice. Lighter shades of grey correspond to higher
density. The left (right) panel shows the evolution of a positive
(negative) $m^*$ state. The dimensionless time is $t$=0, 1, 2, and
3 from the top to bottom. In the plots, the ranges for $x$ and $y$
are from $-30$ to 30. The condensate parameters are $N=50000$,
$g=-0.004$, $V_0=10$, $k_0=3$, $K_3=3.9 \times 10^{-8}$ and the
width of the initial envelope function $w=8$. The units for time,
length and energy are $1/\omega$, $\sqrt{\hbar/(m\omega)}$, and
$\hbar \omega$, respectively, where $\omega$ is the trap frequency
in the axial direction.} \label{fig2}
\end{figure}

\begin{figure}
\includegraphics*[width=7cm,height=5.cm]{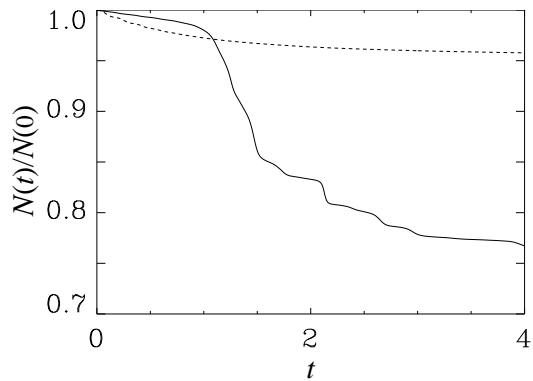}
\caption{Temporal evolution of the number of atoms in the
condensate. The solid (dashed) line is for the positive (negative)
$m^*$ state. Same parameters as in Fig.~\ref{fig2}. } \label{fig3}
\end{figure}

The evolution of the transverse component $\psi_\bot$ of the
condensate wave function is then given by the generalized
Gross-Pitaevskii equation
\begin{eqnarray*}
&&i\hbar \frac{\partial \psi_\bot}{\partial t} \nonumber \\&=&
\left[- \frac{\hbar^2}{2m} \nabla_{\bot}^2 +
V(x,y)+Ng|\psi_\bot|^2-iN^2 K_3 |\psi_\bot|^4 \right] \psi_\bot,
\end{eqnarray*}
where
$$\nabla_{\bot}^2 =
\partial^2/\partial x^2 + \partial^2/\partial y^2.$$
For the initial state of the condensate, we assume a gaussian
transverse wave function of the form
\begin{equation}
\psi_\bot(x,y,t=0) ={\cal N} \,f(x,y)\, \phi_{n{\bf k}}(x,y),
\end{equation}
where $f(x,y)=\exp[-(x^2+y^2)/w^2]$ is a slowly varying Gaussian
envelope function of width $w$, $\phi_{n{\bf k}}$ is a Bloch wave
function of the linear Hamiltonian, and the constant ${\cal N}$
normalizes $\psi_\bot$ to unity. Figs.~\ref{fig2} and \ref{fig3}
illustrate the dynamics of the condensate for two cases, one
corresponding to a negative effective mass and the other to a
positive one. We choose the Bloch wave function $\phi_{n{\bf k}}$
associated with a negative effective mass $m^*$ as
\begin{equation}
\phi_{n{\bf k}} \sim \sin (k_0 x) \,\sin(k_0 y).\label{neg1}
\end{equation}
This corresponds approximately to the Bloch state at the top of
the first energy band, whose full expression is given by
$\phi_{1,s=1}$ of Eq.~(\ref{neg}). The approximate form
(\ref{neg1}) retains only the dominant term in the Fourier series
(\ref{neg}). Its overlap with $\phi_{1,s=1}$ is larger than 99\%
for the parameters of our simulations. Similarly, for the state of
positive effective mass $m^*$ we use
\begin{equation}
\phi_{n{\bf k}} \sim \cos (k_0 x) \,\cos(k_0 y),\label{pos1}
\end{equation}
which corresponds roughly to the Bloch state $\phi_{2,s=1}$ at the
bottom of the second band, see Eq.~(\ref{pos}). Both states
(\ref{neg1}) and (\ref{pos1}) are located at the edge of the first
Brillouin zone, with corresponding drift velocities ${\bf v}_g=0$.
Hence the center of the condensate stays at $x=y=0$ during the
course of its time evolution.

The two columns in Fig.~\ref{fig2} show snapshots of the density
distribution of the condensate for positive (left column) and
negative (right column) effective masses $m^*$. In both cases the
nonlinear interaction strength $g$ is negative, hence in the
absence of the lattice the condensate would be unstable and
subject to collapse. This however can be changed if the effective
mass of the BEC becomes negative, as demonstrated in the right
column. In that case, the condensate behaves as if experiencing a
repulsive two-body interaction, its width increasing in time. This
is in sharp contrast to the situation for a positive effective
mass, in which case the condensate width quickly implodes
\cite{note}. This implosion is characterized by a complex
dynamical behavior, such as the formation of atomic ``bursts,''
observed at later times. (We remark that while an attractive
condensate can expand if it possesses a sufficiently large kinetic
energy, this is not the case of Fig.~\ref{fig2}, where the two
initial states have nearly identical kinetic energies.) The
collapse dynamics depicted here is rather similar to that of an
attractive condensate in a harmonic potential, a system which has
gained much interest recently as a result of the capability of
tuning the scattering length from positive to negative via
Feshbach resonance \cite{jila,saito,santos}. We note however that
the use of the generalized Gross-Pitaevskii equation becomes
questionable in the regime of extreme collapse. This point though
it is important, is not central to our results.

We conclude this section by observing that the number of
condensate atoms decreases in time, due to the presence of
three-body loss. This is particularly evident for the state with
positive effective mass, as shown in Fig.~\ref{fig3}. This is as
expected, since the atomic density is higher in that case.
Finally, we note that as a result of the tensorial character of
the effective mass, it can readily be positive along some
dimensions and negative along others. For example, in two
dimensions a condensate wave function of the form
\[ \phi_{n{\bf k}} \sim \sin (k_0 x)\,\cos (k_0 y), \]
has an effective mass that is negative
along $x$ and positive along $y$-direction. This leads to a
situation where the condensate will expand along $x$ and contract
along $y$, as has been numerically confirmed.

\subsection{Inverted external potential}
\label{gravity}

We mentioned that if a condensate of negative effective mass is
subject to an external potential $U({\bf r})$ that varies slowly
over the lattice period, then this potential will appear inverted
to the condensate. For example, under the effect of gravity, the
center of mass of the condensate will climb up, rather than fall
down. To demonstrate this effect, we solve the one-dimensional
Gross-Pitaevskii equation
\begin{equation}
i\hbar \frac{\partial \psi}{\partial t} = \left[
-\frac{\hbar^2}{2m}\frac{\partial^2}{\partial z^2} + V(z)+U(z)+Ng
|\psi|^2 \right] \psi ,
\end{equation}
where $V(z)=V_0 \cos^2 (k_0 z)$ is the lattice potential and
\[U(z)=G z,\] represents the gravitational potential, $G$ being a
constant. We neglect the three-body loss term, which is not
essential in the present discussion.

\begin{figure}
\includegraphics*[width=7cm,height=10.cm]{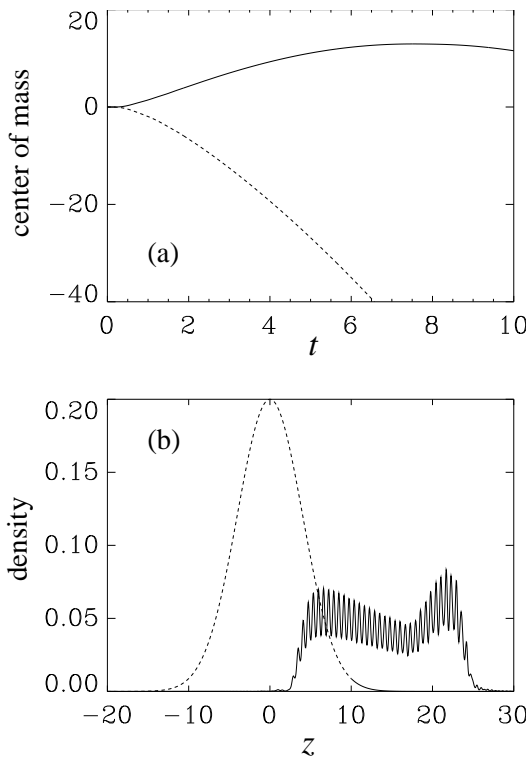}
\caption{(a) Temporal evolution of the condensate center of mass.
(b) Solid line: density distribution of the state with negative
effective mass at the dimensional time $t=7.53$ when the center of
mass of the condensate reaches maximum height. The dashed line
shows the initial Gaussian envelope function for reference. The
parameters used are $V_0=15$, $w=8$, $k_0=5$, $G=0.65$, and
$Ng=10$, in the same dimensionless units as in Fig.~\ref{fig2} is
used. } \label{fig4}
\end{figure}

Again, we assume that the initial condensate wave function is a
broad Gaussian envelope function centered at $z=0$, modulated by a
sinusoidal function. Taking that modulation of the form $\sin (k_0
z)$ yields a state of negative effective mass $m^*$, while $\cos
(k_0 z)$ gives a state of positive effective mass.
Figure~\ref{fig4}(a) illustrates the evolution of the center of
mass of the condensate for these two cases. As expected from the
previous discussion, we see that it initially climbs up for
negative $m^*$ and falls down for positive $m^*$.

As the center of mass of the condensate of negative effective mass
climbs up against the gravitational field, its gravitational
potential energy increases. This increase must be offset by a
decrease in kinetic energy (assuming for the sake of argument that
the nonlinear interaction energy is negligible). Since the initial
condensate possesses a finite kinetic energy, roughly given by
$\hbar^2 k_0^2/(2m)$, conservation of energy therefore sets an
upper limit on how high its center of mass can climb. This is
illustrated in Fig.~\ref{fig4}(a). A good estimate for the maximum
height can be obtained from
\[\frac{\hbar^2 k_0^2}{4m} \approx mG z_{\rm max},\]
where the left hand side is about half of the initial kinetic
energy and the right hand side is the maximum gain in
gravitational energy. For lithium atoms in an optical lattice
formed by laser light of wavelength of 0.5 $\mu$m, one finds that
$z_{\rm max}$ is about 0.4 mm, a rather significant distance in
the context of integrated atom optics \cite{iao}.
Figure~\ref{fig4}(b) compares the condensate density distribution
at the maximum height to its initial envelope profile.

That a maximum height should exist demonstrates the limit of the
effective mass formalism, since a literal interpretation of the
effective mass equation (\ref{gpe*}) would incorrectly lead one to
conclude that in the case of negative effective mass, the
condensate should move against the gravitational field without
bound. The incorrectness of this reasoning lies in the fact that
Eq.~(\ref{gpe*}) neglects all terms coupling different Bloch modes
(see derivation in Appendix~\ref{app1}). What happens in reality
is that as the condensate moves under the combined influence of
the external potential and of its intrinsic nonlinear potential,
other Bloch modes with different quasi-momenta than those
initially excited inevitably become populated. This in turn
changes the effective mass of the system, or even makes this
band-specific concept meaningless. In the present example, the
effective mass changes its sign to positive at the maximum height
as successive energy bands become excited.  In the mean time, the
oscillations in the density profile have a contrast less than one
[see Fig.~\ref{fig4}(b)] which shows that the phase of the wave
function at maximum height becomes spatially continuous
(otherwise, the density would vanish at each phase singularities)
while the initial wave function possesses a series of $\pi$-phase
jumps.

\subsection{Discussion}

We conclude this section with some general comments on the
generation and robustness of condensate states of negative
effective mass. As one can see from Eq. (\ref{neg1}), the
condensate wave function suffers a $\pi$-phase jump every lattice
period. This immediately suggests that these states can be
prepared by the phase-imprinting method that has been previously
successfully implemented to generate dark solitons as in
Ref.~\cite{phase}. However, due to edge effects and inaccuracy in
controlling the intensity of the phase imprinting pulse, this
method may not be fully adequate to generate a series of accurate
$\pi$-jumps as required by Eq.~(\ref{neg1}).

A better choice may therefore consist in first preparing a
condensate with uniform phase in a specific electronic state.
Taking advantage of the internal degrees of freedom of the atoms,
one can then drive the condensate into another internal state with
an optical field of appropriate topological spatial geometry. The
resulting center-of-mass wave function of the condensate having
then the correct phase structure. This method has been
successfully used in the past to create a series of accurate
$\pi$-phase jumps in condensates \cite{pi1,pi2,pi3}.

We conclude this section with a brief discussion of the robustness
of condensate states of negative effective mass. As we have
mentioned, the state (\ref{neg1}) is only an approximation to the
true Bloch state (\ref{neg}). The overlap between these two states
decreases as $h^2$ increases. Furthermore, $|1/m^*|$ decreases
with increasing $h^2$, as the system approaches the tight-binding
regime. On the other hand, $h^2$ cannot be made too small, since
the size of the gap between the first and second energy band is
itself proportional to $h^2$, and a small gap will result in
inter-band transitions to a state of positive effective mass. The
optimal value od $h^2$ also depends on other system parameters,
and as a result we found that $h^2 \sim 2$ turned out to be
optimal for our purpose.

\section{Conclusion}

In summary, we have studied the dynamics of an atomic condensate
in a periodic optical lattice using an effective mass method. We
have obtained an effective equation of motion governing the time
evolution of the envelope of the condensate wave function in which
the periodic external potential appears in the form of an
effective mass, which can be either positive or negative, and a
global drift. Numerical calculation confirmed that this envelope
function approach provides us with useful qualitative insight into
the condensate dynamics. Our study focused on negative effective
masses, in which case the condensate behaves as if it subject to
an inverted nonlinearity and an inverted external potential.

We emphasize that caution must be used when applying the effective
mass concept, which neglects the potentially important coupling
between different Bloch modes. This is particularly so for a
system with negative effective mass moving in an external
potential, as illustrated in Sec.~\ref{gravity}. An ordinary
particle inside an external potential will move towards the
potential minimum, in the mean time gaining kinetic energy. A
particle with negative effective mass, however, will move towards
the potential maximum and in doing so, it looses kinetic energy.
This process cannot go on forever since there is only a finite
amount of kinetic energy for the system to loose. In the process,
the effective mass will eventually change from negative to
positive.

Our study also demonstrates that, as an alternative to the
established Feshbach resonance method, changing the effective mass
of a Bose-Einstein condensate provides us with an additional way
to control the nonlinear atom-atom interaction in a condensate.
This approach should be particularly useful for systems with no
Feshbach resonances at convenient magnetic field strengths, or
when the presence of external magnetic fields is undesirable.
Controlling atom-atom interactions via their effective mass also
provides the possibility to induce anisotropic nonlinear
interactions, due to the tensorial character of the effective mass
in higher dimensions.

Our study here focused on the dynamics of the system. In the
future, it will also be interesting to study its static
properties. For doing so, one needs to provide an additional
confining potential. We must keep in mind that a confining
potential for a negative effective mass state is an {\em
anti-trap}, instead of a trap. This suggests the possibility of
trapping strong-field-seeking states using the magnetic traps
implemented in current BEC experiments.

\begin{acknowledgments}
This work is supported in part by the US Office of Naval Research,
by the National Science Foundation, by the US Army Research
Office, by NASA, and by the Joint Services Optics Program. L. O.
Baksmaty wishes to thank the Horton Foundation. We would also like
to thank Michael Banks for invaluable computer support.
\end{acknowledgments}

\appendix
\section{Derivation of the effective mass equation}
\label{app1}

In this Appendix, we give a derivation of the effective mass
equation~(\ref{gpe*}). For the sake of simplicity, we restrict
ourselves to a one-dimensional system and neglect the nonlinear
terms, which can be added straightforwardly.

We are concerned with the solution of the Schr\"{o}dinger equation
\begin{equation}
i\hbar \frac{\partial \psi(z,t)}{\partial t}= \left[ H_0 + U(z)
\right] \psi(z,t) ,\label{a1}
\end{equation}
where $H_0= {\hat{p}^2}/{2m} + V(z)$ is the Hamiltonian for
periodic potential $V$ and $U$ is some additional potential that
varies on a spatial scale much larger than the lattice period.

\subsection{The ${\bf k} \cdot {\bf p}$ perturbation method}

Associated with each quasi-momentum $\hbar k$ is a set of Bloch
functions $\phi_{n k}$ with energy $E_{nk}$ as defined in
Eq.~(\ref{bloch}). In this section, the power expansion around the
central wave vector $k_0$ of the energy band function $E_{nk}$ is
investigated. In particular, we seek expressions for first and
second derivatives of $E_{nk}$ with respect to $k$. These
expressions will become useful in the derivation of Eq.
(\ref{gpe*}).

The ``${\bf k} \cdot {\bf p}$'' perturbation approach is a
straightforward method to relate the energies and wave functions
at nearby points in the quasi-momentum space. We proceed by
writing the Bloch function $\phi_{n{\bf k}}$ as
\[ \phi_{nk} (z) = e^{ik z}\,
u_{nk}(z),\] where $u_{nk}$ is the cell periodic function. From
Eq.~(\ref{bloch}), we immediately have
\begin{equation}
\left[ \frac{\hat{p}^2}{2m} + V(z) +\frac{\hbar^2 k^2}{2m} +
\frac{\hbar k}{m} \hat{p} \right] u_{nk}= E_{nk} \,u_{nk}.
\label{unk}
\end{equation}
We then perform a power expansion around $k_0$ up to second order,
such that
\begin{eqnarray*}
k &=& k_0 + \delta k ,\\
E_{nk} &=& E_{nk_0} + (\delta k )\,\left( \frac{\partial
E_{nk}}{\partial k}\right)_0 + \frac{1}{2}(\delta k)^2\,
\left(\frac{\partial^2 E_{nk}}{\partial k^2}\right)_0,\\
u_{nk} &=& u_{nk_0} + (\delta k ) \sum_{m \neq n} a_m^{(1)} u_{m
k_0} + (\delta k )^2 \sum_{m \neq n} a_m^{(2)} u_{mk_0},
\end{eqnarray*}
where $(\cdots)_0$ means that the corresponding derivative is
evaluated at $k=k_0$. Next we insert these expansions into
Eq.~(\ref{unk}), and equate the coefficients of different powers
of $\delta k$. For the zeroth order, we find
\[ \left[
\frac{\hat{p}^2}{2m} + V(z) +\frac{\hbar^2 k_0^2}{2m} +
\frac{\hbar k_0}{m} \hat{p} \,\right] u_{nk_0}= E_{nk_0}
\,u_{nk_0},\] which is simply the definition of $u_{nk_0}$. The
first-order term yields
\begin{eqnarray}
&&\left[ \left( \frac{\partial E_{nk}}{\partial k} \right)_0
\right.  - \left. \frac{\hbar^2k_0}{m} + \frac{\hbar}{m}\hat{p}\,
\right]\,u_{nk_0} = \nonumber \\ && \sum_{m \neq n} a_m^{(1)}
(E_{mk_0}-E_{nk_0}) u_{m k_0}. \label{first}
\end{eqnarray}
Taking the dot product of Eq. (\ref{first}) with $u_{nk_0}$ and
integrating over $z$ in the first Brillouin zone, we find
\begin{eqnarray*}
&&\frac{1}{\hbar} \left( \frac{\partial E_{nk}}{\partial
k}\right)_0 = \frac{\hbar k_0}{m} + \frac{1}{m}\, p_{nn}(k_0) \\
&&= \frac{1}{m} \langle \phi_{nk_0} | \hat{p}\,|
\phi_{nk_0}\rangle = v_g,
\end{eqnarray*}
where we have used the orthonormality condition of the Bloch
functions and \[ p_{nl}(k) \equiv \frac{2\pi}{\Omega} \langle
u_{nk}|\hat{p}\,|u_{lk} \rangle.\] One can see that the drift
velocity as defined in Eq.~(\ref{vg}) is just the gradient of the
band energy at $k_0$.

Carrying out a similar procedure with $u_{mk_0}$ ($m \neq n$), we
get from (\ref{first}) the expression for $a_m^{(1)}$ as
\[ a_m^{(1)} = \frac{\hbar}{m} \frac{p_{mn}}{E_{mk_0}-E_{nk_0}} .\]
Analogously, by using the second-order equation, we can find the
second derivative of $E_{nk}$ as
\[ \left( \frac{ \partial^2 E_{nk}}{\partial k^2} \right)_0 =
\frac{\hbar^2}{m} + \frac{\hbar^2}{m^2} \sum_{m \neq n} \frac{2
p_{mn} p_{nm}} {E_{mk_0} - E_{nk_0}}.\] Since $1/m^*= (\partial^2
E_{nk}/\partial k^2)_0/\hbar^2$, we have
\begin{equation}
\frac{1}{m^*} = \frac{1}{m}+\frac{1}{m^2} \sum_{m \neq n} \frac{2
p_{mn} p_{nm}} {E_{mk_0} - E_{nk_0}}. \label{meff}
\end{equation}
\vspace{.6 cm}

\subsection{Derivation of Eq.~(\ref{gpe*})}

We now proceed to derive Eq.~(\ref{gpe*}, adaptingwith slight
modifications from the approach of Ref.\cite{book}. First, we
introduce the Fourier expansion of the envelope function as
\[ f_n(z,t) = \int A_{nk}(t) e^{i(k-k_0)z} \,dk .\]
From Eq.~(\ref{exp}), we have
\begin{equation}
\psi(z,t) = \sum_n \int A_{nk}(t)\,
\chi_{nk}(z)\,e^{-iE_{nk_0}t/\hbar} \,dk,\label{exp1}
\end{equation}
where
\[\chi_{nk}(z) =e^{i(k-k_0)z} \phi_{nk_0}(z) =e^{ikz}
u_{nk_0}(z).\]

After some algebra, one finds the matrix elements of $H_0$ and $U$
as
\begin{eqnarray*}
&&\langle \chi_{nk}|H_0| \chi_{lq} \rangle = \\
&&\delta(q-k) \left\{ \left[
E_{lk_0}+\frac{\hbar^2(q^2-k_0^2)}{2m} \right] \,\delta_{nl}+
\;\frac{\hbar(q-k_0)}{m}p_{nl}(k_0) \right\}, \\
&&\langle \chi_{nk}|U| \chi_{lq} \rangle = \delta_{nl}
\tilde{U}(k-q),
\end{eqnarray*}
where
\[\tilde{U}(k-q) = \frac{1}{2\pi} \int U(z) e^{i(q-k)z}\,dz
.\]
Inserting (\ref{exp1}) into (\ref{a1}), we have
\begin{widetext}
\begin{equation}
i\hbar \frac{\partial A_{nk}(t)}{\partial t} =
\frac{\hbar^2(k^2-k_0^2)}{2m} A_{nk} + \frac{\hbar (k-k_0)}{m}
\sum_l p_{nl}(k_0)\,A_{lk}+  \int dq \, \tilde{U}(k-q) \,A_{nq}.
\label{aa}
\end{equation}
\end{widetext}
The only terms that couple different bands ($l \neq n$) are those
involving the momentum matrix elements $p_{nl}$. As long as these
inter-band couplings are weak enough, we can treat them as weak
perturbations. To lowest order, we have
\[ A_{lk} = \frac{\hbar (k-k_0)}{m} \frac{ p_{ln}(k_0)} {E_{lk_0}-
E_{nk_0}} \,A_{nk}.\] Inserting this expression into Eq.
(\ref{aa}), we have
\begin{equation}
i\hbar \frac{\partial A_{nk}}{\partial t} =\left[ \frac{\hbar^2
(\delta k)^2 }{2m^*}  + \hbar v_g (\delta k) \right] A_{nk} +\int
dq \, \tilde{U}(k-q) \,A_{nq},\nonumber
\end{equation}
where $\delta k= k-k_0$. Performing an inverse Fourier transform
from $A_{nk}(t)$ back to $f_n(z,t)$ and replacing $\delta k$ by
its configuration space representation $-i
\partial /
\partial z$, we finally obtain the one-dimensional version of
Eq.~(\ref{gpe*}). The nonlinear terms can be added to this
equation in a straightforward way.


\begin{references}
\bibitem{loss}The two-body loss term has been omitted since it is
normally negligible compared to the three-body loss term.
\bibitem{note} Note that once the collapse of the condensate
occurs, the envelope function cannot be regarded as slowly varying
any more compared to the lattice period. The effective mass
equation becomes invalid in this case.
\end{references}
\end{document}